\documentclass[aps,showpacs,prc,nofootinbib,floatfix]{revtex4}

\usepackage{graphicx}

\usepackage{latexsym}
\usepackage{hhline}
\usepackage{amssymb}
\usepackage{wasysym}
\usepackage{epsfig}
\usepackage{ulem}

\sloppypar

\newcommand{\be}{\begin{equation}}
\newcommand{\ee}{\end{equation}}
\newcommand{\bea}{\begin{eqnarray}}
\newcommand{\eea}{\end{eqnarray}}
\newcommand{\ba}{\begin{array}}
\newcommand{\ea}{\end{array}}
\newcommand{\bi}{\begin{itemize}}
\newcommand{\ei}{\end{itemize}}

\newcommand{\refe}[1]{(\ref{#1})}

\newcommand{\foh}{\frac{1}{2}}

\newcommand{\fth}{\frac{3}{2}}
\newcommand{\ffh}{\frac{5}{2}}

\begin{document}

\title{Eta-meson production in the resonance energy region. 
\footnote{Supported by Transregio SFB/TR16, project B.7}}

\author{V. Shklyar}
\email{shklyar@theo.physik.uni-giessen.de}
\author{H. Lenske} 
\author{U. Mosel} 
\affiliation{Institut f\"ur Theoretische Physik, Universit\"at Giessen, D-35392
Giessen, Germany}

\begin{abstract}
We perform an updated coupled-channel analysis of  eta-meson production including all recent
 photoproduction data on the proton. The dip observed in the differential 
cross sections at c.m. energies W=1.68 GeV is explained by destructive interference between 
the $S_{11}(1535)$
and $S_{11}(1560)$ states.
 The effect from $P_{11}(1710)$ is found to be small but still important to reproduce the
correct shape of the differential cross section.
 For the $\pi^- N \to \eta N$  scattering we suggest a reaction
mechanism in terms of the $S_{11}(1535)$,  $S_{11}(1560)$, and $P_{11}(1710)$ states. 
Our conclusion on the importance of the  $S_{11}(1535)$,  $S_{11}(1560)$, and $P_{11}(1710)$ 
resonances in  the eta-production reactions  is in line  with our previous results.
No strong indication for a narrow state with a width of 15~MeV and the mass of 1680 MeV is found in  the
analysis.   $\eta N$ scattering length is extracted and discussed.

\end{abstract}


\maketitle

\section{Introduction}

The discovery of  nucleon resonances in the first pion-nucleon scattering experiments 
provided  first indications for a complicated  intrinsic structure of the nucleon.
With establishing   the  quark picture of hadrons  and developments of the 
constituent quark models the interest in the study of the nucleon excitation spectra was renewed. 
The major question was the number of the excited states and their properties. This problem was attacked 
both experimentally and theoretically. 
On the theory side constituent quark (CQM) models, lattice QCD and Dyson-Schwinger  approaches 
have been developed  to describe and predict the nucleon  resonance  spectra
(see e.g. \cite{Aznauryan:2009da} for a review).
The main problem remains, however, a serious disagreement between the theoretical calculations and 
the experimentally  observed baryon spectra. This concerns both the number and the properties of excited states. 

On the experimental side  pion-induced reactions   have   been studied 
to establish resonance spectra.  However, due to  difficulties in detecting neutral 
particles 
most  experiments  were limited to  pion-nucleon elastic scattering
with charged particles in the final state.
Being the lightest non-strange particle next to the pion the  $\eta$-meson  also becomes  an 
interesting probe to study nucleon excitations. 
A few experiments have been made in the past  to investigate  $\eta$-production. 
The first near-threshold  measurements  \cite{Jones:1966zza,Richards:1970cy,Bulos:1970zk} demonstrated
that the reaction proceeds through a strong  $S$-wave   
resonance excitation which  was later identified with  $S_{11}(1535)$. 
An extensive  study of the $ \pi^- p \to\eta n$ reaction above  W$>$1.7 GeV  has been  
made in \cite{ Brown:1979ii,Baker:1979aw}. Both  differential cross section and asymmetry 
data have been obtained. However,  due to  possible  problems  with the  energy-momentum
calibration \cite{Clajus:1992dh} the use of these   data might lead to   wrong conclusions on the 
reaction mechanism. Note that these problems 
are  present  not only in the $\eta$-measurements \cite{ Brown:1979ii,Baker:1979aw} 
but also in the charge-exchange data obtained  in  the same  experiment.

Presently the  development of the high-duty electron facilities (ELSA, JLAB, MAMI, SPring) offers 
 new possibilities to study the $\eta$-photoproduction  both on the proton ($\eta p$) and on the  
neutron ($\eta n$). 
The first  measurement of the $\eta$-photoproduction on the neutron reported  an indication 
for a  resonance-like structure in the reaction cross section at 
W=1.68 GeV \cite{Kuznetsov:2006, Kuznetsov:2006kt}.
Independent experimental studies   \cite{Jaegle:2008ux,Jaegle:2011sw} confirmed the 
existence of this effect   in the $\gamma n$  reaction. This phenomenon was 
predicted  in \cite{Azimov:2005jj} as a  signal from a narrow state - a possible 
non-strange partner of the pentaquark \cite{Diakonov:1997mm}. Another
explanation has been suggested in \cite{Shklyar:2006xw} where the   observed effect was  
described  by the contributions  from  the $S_{11}(1650)$ or  $P_{11}(1710)$ states.  
Due to the lack  of  knowledge of the $S_{11}(1650)$ and $P_{11}(1710)$
resonance couplings to  $\gamma n$ a clean  separation of the relative contributions from
these states is difficult.  The general conclusion made in \cite{Shklyar:2006xw}
 is that  both states might be good candidates to explain the 
observed structure. 

 By fitting to the $\eta n$ cross sections and beam asymmetry the Bonn-Gatchina group
provided an  explanation \cite{Anisovich:2008wd} for the second peak in terms of the  $S_{11}(1650)$ state.
 Another contribution to the field has been made by the  authors of \cite{Doring:2009qr}.
There the peak in  the $\sigma_p/\sigma_n$ cross section ratio  was explained by a cusp
effect from   the $K\Sigma$ and $K\Lambda$ rescattering channel. All these studies  
have been done assuming
scattering on a quasi-free nucleon. At the same time a realistic  analysis of 
meson photoproduction on the quasi-free neutron  should include the nucleon-nucleon and meson-nucleon  
correlations (FSI-effect) which were shown to be very important \cite{Tarasov:2011ec} and take into 
account corresponding experimental cuts applied by the extraction of the quasi-free neutron 
data from  $\gamma D$-scattering. The later issue  might be crucial for the  unambiguous identification 
of the narrow resonance contribution  as discussed in  \cite{MartinezTorres:2010zzb}.

If it is granted that the signal  observed in the $\gamma n$ scattering  
\cite{Kuznetsov:2006, Kuznetsov:2006kt,Jaegle:2008ux,Jaegle:2011sw}
 is due to the narrow (exotic) state one may expect to observe a similar effect  in other eta-production 
reactions at the  same  energies, e.g in   gamma-proton scattering.
The experimental investigations of the 
$\eta$-production on the proton  made by the  CLAS, GRAAL, and  CB-ELSA/TAPS collaborations 
\cite{Dugger:2002ft,Crede:2009zzb,Bartholomy:2007zz,Bartalini:2007fg} have found  an  indication of the dip 
structure  around W=1.68 GeV in  the differential cross section but not a resonance-like structure. 
This effect was also accompanied by the change in the angular distribution of the differential cross section. 
However, despite of  extensive theoretical studies of the $\eta$ -production 
 the reaction mechanism is still under discussion 
\cite{PhysRevC.84.045207,Chiang:2002vq,Bartholomy:2007zz,Feuster:1998b,An:2011sb,Anisovich:2011ka,Shyam:2008fr,
Nakayama:2008tg,Choi:2007gy,Zhong:2007fx,Fix:2007st,Gasparyan:2003fp,Ruic:2011wf}.
 
Recently the $\eta$-photoproduction on the proton has been measured with high-precision 
by the Crystal Ball collaboration at MAMI \cite{McNicoll:2010qk}. 
These high-resolution data  provides a new step forward 
in understanding  the reaction dynamics and in the search for a signal from the 
'weak' resonance states.  
The main result  reported in \cite{McNicoll:2010qk} is a very clean signal of a dip 
structure around W=1.68 GeV. It is interesting to note that the  old measurements of the 
$\pi N \to \eta N$ reaction \cite{Richards:1970cy} also give an indication for the second structure in the
differential cross section at W=1.7 GeV. This raises a question whether the 
dip reported in the $\eta p$ reaction, the resonance-like signal observed in $\eta n$ and the  possible
structure in the $\pi N \to \eta N $ cross section are originating from the same degrees of freedom
or not. The second question is whether one of these  phenomena 
can be attributed to the signal from  a narrow (exotic) resonance state as discussed 
 in \cite{Azimov2,Azimov:2005jj,Azimov:2005hv}.

In our previous  coupled-channel PWA study  \cite{Shklyar:2006xw}
 we proposed an explanation of  the possible  dip in the $\eta$-proton cross section  in terms of
the destructive interference of  the  $S_{11}(1535)$ and $S_{11}(1650)$ states. 
The result was  based on the $\eta p$ photoproduction data taken before 2006  \cite{Dugger:2002ft,Crede:2003ax}.
The aim of the present study is to extend our previous coupled-channel analysis of the $\gamma p\to \eta p$ 
reaction by including the data from the high-precision  measurements  \cite{McNicoll:2010qk}. 
The main question is whether the $\eta p$ reaction dynamics can be understood in terms of the established 
resonance states. We emphasize that for  reliable  identification of 
the resonance contributions the calculations should maintain unitarity.  Another complication comes 
from the fact that the most contributions to the resonance self-energy (total decay width) is driven by its hadronic 
couplings. Therefore the  analysis  of the photoproduction data requires the knowledge of the 
hadronic transition amplitudes. Hence the simultaneous analysis of all open channels (both hadronic 
and electromagnetic ) is inevitable for the identification of the resonances and extraction of their properties.  
In the present study we concentrate on the combined description of the $(\gamma/\pi) p \to \eta p$ scattering
taking also the $(\gamma/\pi) \to \pi N$, $2\pi N$, $\omega N$, $K \Lambda$ channels  into account.  
The results on the $\eta n$ reaction will be reported elsewhere. 

First, we corroborate  our previous findings \cite{Penner:2002a,Penner:2002b,Shklyar:2006xw} 
where the important contributions from the $S_{11}(1535)$, $S_{11}(1650)$ and $P_{11}(1710)$ 
resonances to the $\pi N \to \eta N$ reaction have been found. 
The major effect comes from the $S_{11}$ and $P_{11}$ partial waves.
The interference  between  the $S_{11}(1535)$ and $S_{11}(1650)$ states 
produces a dip in the $S_{11}$ amplitude. The $P_{11}$ amplitude is 
influenced by the contributions from the $P_{11}(1710)$ state. 
The  interference between the $S_{11}$ and $P_{11}$ partial waves leads to the 
forward peak in the differential cross section around W=1.7 GeV. 
We stress that the interference between two nearby states also includes rescattering and  coupled-channel 
effects which are hard to simulate by the simple sum of two Breit-Wigner forms.

We also confirm our previous finding that the interference between $S_{11}(1535)$ and  $S_{11}(1650)$ 
is responsible for the dip seen in the $\eta p$ data. The effect from the $\omega N$ threshold is found to be
relatively small which is also in line with the  conclusion of \cite{Shklyar:2006xw}.   Opposite to 
\cite{Anisovich:2011ka} we do not find any strong indications for a narrow state  in the Crystal Ball/Taps 
data around W=1.68 GeV.
We have also checked our results for 
 the $\eta p$ reaction above W=2 GeV where a number of new experimental data are available. Note that we do not
use Reggezied  $t-$channel exchange but include all $t$-channel contributions consistently  into 
our unitarization procedure.  Because of the  normalization 
problem \cite{Sibirtsev:2010yj,Dey:2011rh} between the  CLAS \cite{Williams:2009yj} and 
the CB-ELSA \cite{Crede:2009zzb} datasets  
 the simultaneous description of these data is not possible. 
Above W=2 GeV our calculations are found to be in closer agreement with the
CLAS measurements \cite{Williams:2009yj}. The CB-ELSA data \cite{Crede:2009zzb} 
demonstrates a step rise around W=1.925 GeV for the scattering angles $\cos\theta=0.85...0.95$. 
It is not  clear whether this phenomenon could be related to a threshold effect  (e.g. $\phi N$, $a_0(980) N$, 
$f_0(980)$,  or $\eta' N$) or attributed to other reaction mechanisms. 

We conclude   that  further progress in understanding of the $\eta$-meson production dynamics 
would be hardly possible without new measurements of the $\pi N\to \eta N$ reaction.

\section{Database}
\label{data}
Here we present a short overview of the experimental database relevant for the present calculations. 
The details  on the 
$K\Lambda$, $K\Sigma$, $\omega N$ channels will be given elsewhere.

\mbox{ $\pi N\to\eta N$}:
The thorough overview of the  $\pi N\to\eta N$  experimental  data  (except the 
recently published   Crystal Ball
measurements \cite{Prakhov}), is given in \cite{Clajus:1992dh}.
As already mentioned in Introduction only few  measurements of the $\eta$-production have been made 
with  pion beams: except  for \cite{Danburg:1971} where the eta-meson was produced in $\pi^+ D$ 
collisions, all the data have been taken from  the   $\pi^- p$ scattering  
\cite{Prakhov,Baker:1979aw,Brown:1979ii,Bulos:1970zk,Richards:1970cy,Jones:1966zza,Debenham,Deinet:1969cd}.
Unfortunately due to  numerous problems with the experimental data from  \cite{Baker:1979aw,Brown:1979ii} 
(see discussion  in  \cite{Clajus:1992dh} and references therein) the use of these measurements in the 
analysis might lead to  wrong conclusions for the reaction mechanism. 
Therefore, opposite to \cite{Shrestha:2012va} we do not include these data in the analysis.
Another measurement available above W=1.65 GeV is the   data from  Richards et al 
\cite{Richards:1970cy}. In the first resonance energy region this cross section
tends to be lower than results from other experiments.  
Since the old measurements quote only statistical uncertainties the reason for these 
differences is unclear.  In their study the authors of \cite{Batinic:1995} added  systematical 
errors to all differential cross sections. We do not follow this procedure and include only quoted 
uncertainties in the analysis.

$\gamma p\to\eta p$: 
a number of experimental studies have been performed in the resonance energy region \cite{McNicoll:2010qk, 
Williams:2009yj,GRAAL:2002, Dugger:2002ft,Krusche:1995nv,Bartalini:2007fg,Nakabayashi:2006ut,Crede:2009zzb,Crede:2003ax,
Bartholomy:2007zz,Elsner:2007hm,Bock:1998rk,Ajaka:1998zi}. Most of these measurement are differential cross sections.
The target asymmetry has been studied in \cite{Bock:1998rk}.  It has been observed  that close to the 
$\eta N$ production threshold the asymmetry changes the sign at moderate  scattering angles. 
The previous calculations of the
Giessen Model \cite{Penner:2002a,Shklyar:2006xw} and the Mainz group  \cite{PhysRevC.60.035210} could not explain 
this feature. The description of this data would   require an unexpected phase shift between the $S_{11}$ and $D_{13}$ 
resonances as noted in \cite{PhysRevC.60.035210}. One may hope that the upcoming new measurements of the target 
asymmetry at the ELSA facility will solve this puzzle \cite{Hartmann:2011uv}. 

For the beam asymmetry we use the recent data from the  GRAAL \cite{Bartalini:2007fg} 
and CB-ELSA/TAPS \cite{Elsner:2007hm}  collaborations  which cover the energy region up to W=1.91 GeV.
For the differential cross section we use the recent high-quality Crystal Ball data \cite{McNicoll:2010qk}.
Above W=1.89 GeV  our calculations are constrained by the  amalgamated data set from experiments 
\cite{Williams:2009yj,Crede:2009zzb,Crede:2003ax,Bartholomy:2007zz}.
Since the experimental uncertainties of the data  \cite{Williams:2009yj,Crede:2009zzb,Crede:2003ax,Bartholomy:2007zz}
are much larger than those in \cite{McNicoll:2010qk} we reduce them by factor of 2.

In the $(\pi/\gamma) N \to \pi N$ channels our calculations  are constrained by the single-energy solutions 
from the GWU (former SAID) analysis \cite{Workman:2011vb,Arndt:2006bf,Arndt:2008zz}.
For the $\pi N\to 2\pi N$ transitions we follow the procedure described in 
\cite{Penner:2002a,Penner:2002b,Feuster:1998a,Feuster:1998b}. We continue to parameterize the $2\pi N$ channel 
in terms of the effective $\zeta N$ state, where $\zeta$ is an isovector scalar meson of two pion mass:
$m_\zeta=2m_\pi$. The final $\zeta N$ state is only  allowed to couple to nucleon resonances. Therefore the decay
$N^*\to\zeta N$ stands for the sum of transitions $N^*\to \Delta \pi$, $\sigma  N$, $\rho  N$ etc.
This procedure allows for the good description of the  $\pi N \to 2\pi N$  partial wave cross sections 
extracted in \cite{Manley:1984}. However of case of the $\gamma p\to 2\pi N$ the same agreement 
cannot be expected. This is because of the  enhanced role of the background contributions 
(due to e.g. the  contact $\gamma\rho NN$ interaction in the  $\gamma N\to \rho N$ transitions).
After fixing the database a $\chi^2$ minimization is performed to fix the  model parameters. 

\section{Giessen Model}
\label{model}
Here we briefly outline the main ingredients of the model. More details can be found in 
\cite{Feuster:1998a,Feuster:1998b,Penner:2002a,Penner:2002b,shklyar:2004a,shklyar:2005c}. The Bethe-Salpeter equation 
is solved in the $K$-matrix approximation to obtain  multi-channel  scattering $T$-matrix:
\bea
T(\sqrt{s},p,p') = K(\sqrt{s}, p, p') + \int \frac{d^4q}{(2\pi)^4} 
K(\sqrt{s},p,q)G_{BS}(\sqrt{s},q)T(\sqrt{s},q,p'),
\label{bse}
\eea
where $p$ ($k$) and $p'$ ($k'$) are the incoming and outgoing baryon (meson) four-momenta,
$T(\sqrt{s},p,p')$ is a coupled-channel scattering amplitude, $G_{BS}$ is a meson-nucleon
propagator and $K(\sqrt{s}, p, p')$ is an interaction kernel. The quantities $T(\sqrt{s},p,p')$, $G_{BS}$, 
and  $K(\sqrt{s}, p, p')$ are in fact multidimensional matrices  where the elements of the matrix 
stand for the different scattering reactions. 

To solve the coupled-channel scattering problem with a large number of inelastic  channels, 
we apply the so-called K-matrix approximation by neglecting  the real part of the BSE propagator
$G_{BS}$. After the integration over the relative
energy, Eq. \refe{bse} reduces to
\bea
T^{\lambda_f\lambda_i}_{fi} = K^{\lambda_f\lambda_i}_{fi} + i\int d\Omega_n 
\sum_{n}\sum_{\lambda_n} T^{\lambda_f\lambda_n}_{fn}K^{\lambda_n\lambda_i}_{ni},
\label{GBS2}
\eea
where $T_{fi}$ is a scattering matrix and $\lambda_i$($\lambda_f$) stands for the quantum 
numbers of  initial(final) states  
\mbox{$f,i,n =$ $ \gamma N$}, $\pi N$, $2\pi N$, $\eta N$, $\omega N$, $K\Lambda$, $K\Sigma$. 
Using the partial-wave decomposition of $T$, $K$ in terms of Wigner  d-functions 
the angular integration can be easily carried out and 
the equation is further  simplified to the algebraic form
\bea
T^{J\pm,I}_{fi} = \left[\frac{ K^{J\pm,I}}{1-iK^{J\pm,I}}\right]_{fi}.
\label{GBS3}
\eea

The validity of this approximation
was demonstrated by Pearce and Jennings in \cite{Pearce:1990uj} by studying different approximations
to the BSE for  $\pi N$ scattering. Considering different BSE propagators 
they concluded that an important feature of the reduced intermediate two particle
propagator is the on-shell part of $G_{BS}$. It has been argued
that there is no much difference between physical parameters obtained using the $K$-matrix 
approximation and other schemes. It has also been shown in \cite{Goudsmit:1993cp,Oset:1997it} 
that  for   $\pi N$ and $\bar K N$ scattering the  main effect from the off-shell part  is
a renormalization of the couplings and the masses.

Due to the smallness of the electromagnetic coupling the dominant contributions to the self energy stem
from the hadronic part.   Therefore we treat the photoproduction reactions perturbatively. 
This is equivalent to neglecting $\gamma N$ in the sum over intermediate 
states $n$ in Eq.~\refe{GBS2}. Thus, for a photoproduction process the equation  \refe{GBS3} can be 
rewritten as follows \cite{Penner:2002b,Feuster:1998b}
\bea
T^{J\pm,I}_{f\gamma} = K^{J\pm,I}_{f\gamma} + i \sum_{n} T^{J\pm,I}_{f n} K^{J\pm,I}_{n \gamma}, 
\label{photo}
\eea
where the  summation in  Eq.\refe{photo} is done  over all hadronic intermediate states. Here the  matrix $T^{J\pm,I}_{f n}$
stems only from the  hadronic transitions:  indices $f$ and $n$ run over $\pi N$,    $2\pi N$, $\eta N$, $K\Lambda$,          
$K\Sigma$, $\omega N$ channels. The  sum in  Eq. \refe{photo} reflects  the importance of the hadronic part
of the transition amplitude in the description of photoproduction reactions. In other words, the amplitudes 
for the $\pi N \to \pi N$,  $\eta N$, $\omega N$ etc. transitions should always be included in the calculation
of the photoproduction amplitudes.

\subsection{Interaction kernel and resonance parameters}
\begin{figure}
  \begin{center}
    \includegraphics[width=14cm]{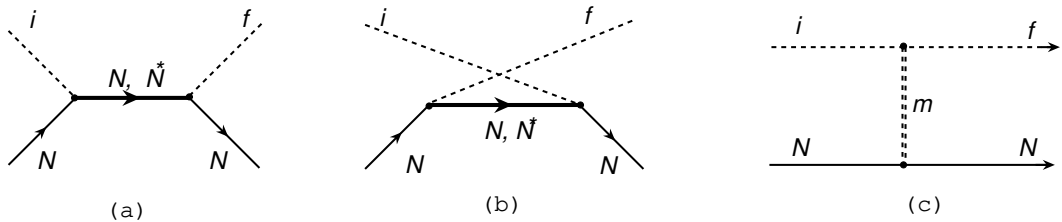}
    \caption{$s$-,$u$-, and $t$- channel contributions to the interaction potential. $i$ and $f$
      stand for the initial and final    $ \gamma N$, $\pi N$, $2\pi N$, $\eta N$, 
       $\omega N$, $K\Lambda$, $K\Sigma$ states. $m$ denotes intermediate  $t$-channel meson. 
      \label{diag}}
  \end{center}
\end{figure}

Here we present the main ingredients of the interaction kernel to the BSE Eq.\refe{bse} relevant for 
$\eta$-production. More details on other reactions can be found in 
\cite{shklyar:2004a,Penner:2002a,Penner:2002b,Feuster:1998a,Feuster:1998b,shklyar:2004b}.
The interaction potential ($K$-matrix) of the BSE is built up as a sum of 
$s$-, $u$-, and $t$-channel contributions corresponding to the tree level Feynman diagrams 
shown in Fig.~\refe{diag}.
In the isospin $I=\foh$ channel we checked for the contributions from the 
$S_{11}(1535)$, $S_{11}(1650)$, $P_{11}(1440)$, $P_{11}(1710)$, $P_{13}(1720)$, $P_{13}(1900)$, $D_{13}(1520)$ 
$D_{13}(1900)$, $D_{15}(1675)$, $F_{15}(1680)$ , $F_{15}(2000)$ resonances.
The resonance and background contributions are consistently generated from the same effective interaction.
 The Lagrangian densities are given in 
\cite{shklyar:2004a,Penner:2002a,Penner:2002b,Feuster:1998a,Feuster:1998b,shklyar:2004b} and respect the chiral symmetry
in low-energy regime.
The properties of the  $t$-channel mesons important for $\eta$ production  are 
given in Table~\ref{tchannel}.
 \begin{table}
  \begin{center}
    \begin{tabular}
      {l|c|r|r|c}
      \hhline{=====}
      & mass [GeV] & $J^P$ & $I$ & final state \\ 
      \hhline{=====}
      $\omega$ & 0.783   & $1^-$ & $0$    & $(\gamma,\eta)$ \\
      $\rho$   & 0.769   & $1^-$ & $1$    & $(\pi,\eta)(\gamma,\eta)$ \\
      $a_0$    & 0.983   & $0^+$ & $1$    & $(\pi,\eta)$ \\
      $\phi$   & 1.02    & $1^-$ & $0$    & $(\gamma,\eta)$ \\
      \hhline{=====}
    \end{tabular}
  \end{center}
  \caption{Properties of mesons  which give contributions
    to the $\eta N$ final state via the $t$-channel exchange. The notation $(\gamma,\eta)$ means 
$\gamma N\to \eta N$ etc.  
    \label{tchannel}} 
\end{table}
Using the interaction Lagrangians and values of the corresponding meson
decay widths taken from  the PDG \cite{pdg} the following coupling constants are obtained: 
\bea
\ba{lcrclcr}
g_{a_0 \eta \pi}       &=& -2.100 \; , & & g_{\omega \eta \gamma} &=& -0.27\; , \\
g_{\rho \eta \gamma}   &=& -0.64 \; , & & g_{\phi \eta \gamma}   &=& -0.385 \; . \\
\ea
\label{mesdeccons}
\eea

All other coupling constants were allowed to be varied during the fit.
The obtained values are given in Table~\ref{BornCoupling}. For the $\eta NN$ interaction  we use 
pseudoscalar coupling , which has been also utilized in our previous studies
\cite{Feuster:1998b,Feuster:1998a,Penner:2002a,Penner:2002b,Shklyar:2006xw}.  
The derived $g_{\eta NN}$ constant is found to be small which is in line with 
our previous results \cite{Shklyar:2006xw,Penner:2002b}. To check the dependence of 
our results  on the choice of the $\eta NN$ interaction we have also performed 
calculations with the pseudovector coupling. However also in the latter case only a  
small  $g_{\eta NN}$ coupling constant  has been found.
 
Since the  PDG gives only 
the upper limit for the decay branching ratio R$(\rho\to\pi\eta) <6\times 10^{-3}$ we allowed this constant to 
be varied during fit. However due to  lack of experimental constraints this coupling cannot 
be fully fixed in the present calculation. We find a small overall contribution from the $t$-channel 
$\rho$-meson exchange to the $\pi^- p \to\eta n$ reaction. 
The $g_{\phi NN }$ coupling is calculated from $g_{\omega NN}$ 
using the relation 
$$
\frac{g_{\phi NN }}{g_{\omega NN }} = -\tan\Delta \theta_{\phi/\omega},
$$
where $ \Delta\theta_{\phi/\omega}$ is a deviation from the ideal $\phi$-$\omega$ mixing angle.
Taking $ \Delta\theta_{\phi/\omega}=3.7^0$ from  \cite{pdg} one gets for 
the ratio $g_{\phi NN }/g_{\omega NN } \approx -1/15$. Using this value  a very small contribution
from the $t$- channel $\phi$-meson exchange to the $\eta$-photoproduction has been found. 

\begin{table}
  \begin{tabular}{cccccccc}
 \hline
 \hline
 $g_{\pi NN}$   &  12.85 & $g_{\rho\eta\pi}$ & 0.133 & $g_{\rho NN}$    &   4.98 & $\kappa_\rho$    &   2.18 \\
 $g_{\eta NN}$  &   0.31 & $g_{a_0 NN}     $ & -44.37 & $g_{\omega NN}$  &   7.23 & $\kappa_\omega$  &  -1.50 \\
 \hline
 \hline
 \end{tabular}\\
\caption{Nucleon and $t$-channel couplings obtained in the present study.\label{BornCoupling}}
\end{table}

To take into account  the finite size  of mesons and baryons each  vertex is 
dressed by a corresponding form factor:
\bea
F_p (q^2,m^2) &=& \frac{\Lambda^4}{\Lambda^4 +(q^2-m^2)^2},
\label{formfact} 
\eea
where   $q$ is a c.m. four-momentum of an intermediate particle and $\Lambda$ is a cutoff parameter.
 The  cutoffs $\Lambda$ in Eq.~\refe{formfact} are treated as  free parameters being varied 
during the  calculation. However, we keep the same cutoffs in all channels for a 
given  resonance
spin  $J$ : $\Lambda^{J}_{\pi N}=\Lambda^{J}_{\pi\pi N}=\Lambda^{J}_{\eta N}=...$ etc., 
($J=\foh,~ \fth,~ \ffh$). This significantly reduces the number of free parameters; i.e. for all 
spin-$\ffh$ resonances there is  only one cutoff $\Lambda=\Lambda_{\ffh}$ for all decay channels.
However for the photoproduction reactions we use different cutoffs at the  $s$- and $u$-channel 
electromagnetic vertices. All values are given in Table  \ref{tabcutoff}.  Except for the  spin-$\fth$ states, 
the $s$- and $u$-channel cutoffs  almost coincide.
\begin{table}[t]
  \begin{center}
    \begin{tabular}
      {c|c|c|c|c|c|c|c}
      \hhline{========}
      $\Lambda_N$ [GeV] & 
      $\Lambda_\foh^h$ [GeV] & 
      $\Lambda_\fth^h$ [GeV] & 
      $\Lambda_\ffh^h$ [GeV] & 
      $\Lambda_\foh^\gamma$ [GeV]  & 
      $\Lambda_\fth^\gamma$ [GeV] & 
      $\Lambda_\ffh^\gamma$ [GeV] & 
      $\Lambda_t^{h,\gamma}$[GeV]  \\
      \hhline{========}
  0.952 & 3.0 & 0.97  & 1.13 &  1.69~(1.69)  &  4.20~(2.9)  &  1.17~(1.25)  &  0.7 \\
       \hhline{========}
    \end{tabular}
  \end{center}
  \caption{Cutoff values  for the form factors. 
    The lower index denotes an intermediate particle, i.e. 
    $N$: nucleon, $\foh$: spin-$\foh$ resonance, $\fth$: spin-$\fth$, $\ffh$: 
    spin-$\ffh$ resonance, $t$: $t$-channel meson. The upper index $h$($\gamma$)
    denotes whether  the value is applied to a hadronic or electromagnetic vertex.
    The cutoff values used at electromagnetic $u$-channel vertices are given in brackets.
    \label{tabcutoff}}
\end{table}

The use of  vertex form factors requires  special care for maintaining  
the current conservation  when the Born contributions to photoproduction reactions 
are considered. Since the resonance and intermediate meson vertices are  constructed from 
gauge invariant Lagrangians  they can be independently  multiplied by the  corresponding form 
factors.  For the nucleon contributions to  meson photoproduction  we apply the suggestion of Davidson 
and Workman  \cite{Davidson:2001rk} and use the crossing symmetric common form factor:
\bea
\tilde F(s,u,t) = F(s) + F(u) + F(t) - F(s)F(u) - F(s)F(t)  - F(u)F(t) +  F(s)F(u)F(t).
\label{formfact2}
\eea

The extracted resonance parameters given in Table~\ref{res_param}  are  very close to the values 
deduced in our previous calculations \cite{shklyar:2004b,Shklyar:2006xw} which  indicates the  stability 
of the obtained solution.   However some values changed upon inclusion of  the new MAMI data
\cite{McNicoll:2010qk}. The total width of $S_{11}(1650)$ tends to be larger than that deduced in 
our previous calculations \cite{shklyar:2004b}. The helicity amplitude is also modified  but still is
in  good agreement with the parameter range provided by PDG \cite{pdg}. The opposite effect is found 
for the $P_{11}(1710)$ state where the total width is reduced once  the data of \cite{McNicoll:2010qk} are included. 
The remaining resonance parameters are only slightly modified as compared to our previous results.

The mass and width of the Roper resonance is found to be larger than deduced
in other analyses  \cite{pdg}. However the authors of \cite{Vrana:2000} 
give $490\pm120$ MeV for the total width. The large decay width $545\pm 170$\,MeV has also  been  
deduced by Cutkosky and Wang \cite{Cutkosky:1990}. Note that properties of this state are strongly 
influenced by its decay into the $2\pi N$ final state. Arndt et al \cite{Arndt:1990bp} found a second  
pole structure for the Roper resonance  which might be attributed to the coupling to the $\pi \Delta$ subchannel. 
Since we use a simplified prescription for the $2\pi N$ reaction this effect cannot be  properly described 
in the present calculations.  

The recent   GWU(SAID) study of the $\pi N$ data shows  no evidence for the  $P_{11}(1710)$ 
resonance. An indirect indication for the existence of this  state can be concluded from the analysis
of the $\pi N$ inelasticity and $2\pi N$ cross section in the $P_{11}$ partial wave, 
see discussion in Section~\ref{pe}. We find a  small coupling of this resonance to the $\pi N$
final state. Since a clear signal from this state is not seen in the recent GWU solution, the 
determination of the total width turns out to be difficult. In our calculations we assume that this 
resonance  has a large decay branching ratio to the $\eta N$. However the quality of 
the $\pi^- p \to \eta n$ data does not allow for an unambiguous determination of the  properties of this 
state.  

The mass and width of the $D_{13}(1520)$ is more close to the values obtained by Arndt et al
\cite{Arndt:1995ak}: $1516\pm10$ MeV and $106\pm4$ MeV respectively. 
It is interesting to note that the mass of this resonance deduced from the pion
photoproduction tends to be  10 MeV lower that the values derived from the pion-induced
reactions \cite{pdg}. The second $D_{13}(1900)$ has a  very large decay width. 
We associate this state with   $D_{13}(2080)$ as suggested  in PDG.
This resonance is rated   with two stars and its existence is still under discussion. In our updated 
coupled-channel calculation  of the $\omega$-production \cite{shklyar:2004b} a large $\omega N$ and 
$2\pi N$ decay branching ratios have been obtained.

The properties of other resonances are very close to the values given in PDG. 
Except for  $S_{1}(1535)$ and $P_{11}(1710)$ we find only small resonance couplings to $\eta N$ which is 
in accordance  with our previous conclusions.  One needs to stress that the smallness of the resonance coupling does not
necessarily mean that the contribution from the state is negligible. The $S_{11}(1650)$ state 
produces for example a sizable effect in the eta-production due to overlapping with  $S_{11}(1535)$. 
Another example is  the effect from the $D_{13}(1520)$ state in $\eta$-photoproduction on the proton.
Here  the smallness of the $\eta N$ branching ratio is compensated by the strong  electromagnetic coupling
of this resonance. 
Therefore the effect from this state could be seen in the $E_{2-}$ and $M_{2-}$ multipoles, see Section~\ref{multipoles}.
However in most cases the resonance contributions with small
branching ratios to the eta
are hard to resolve unambiguously. 

\subsection{Pole parameters}
 
It is interesting to compare the poles positions and elastic residues with the  results from other studies, see
Table\,\ref{poles}. The calculated pole masses are very close  to the values  obtained in other
analyses, see \cite{pdg}. The  agreement between  imaginary parts and elastics residues is also good, 
though some  differences exist  between  the present values   and the results from other groups.

 For the $S_{11}(1535)$ state we obtain a smaller elastic residue (for  definition of $|R|$ see \cite{pdg})
$|R|=15$\,MeV which is almost identical to the result of the GWU group $|R|$=16\,MeV \cite{Arndt:2006bf}. 
Both values seem to be out of the range given in PDG \cite{pdg} 50$\pm$20 MeV. 
It is interesting to note that the elastic residue  from \cite{Arndt:2006bf} is included into 
the estimation made in \cite{pdg} but still  does not fit to the provided range. The value  
$\Gamma_{\rm pole}=89$ MeV for the $S_{11}(1650)$ state is also comparable with the result 
from \cite{Arndt:2006bf}: $\Gamma_{\rm pole}=80$ MeV which are again less than the lower 
bound given in  \cite{pdg}.

Though the derived pole mass of $P_{11}(1440)$ is very close to  the values deduced in  other calculations 
we obtain a significantly larger pole width.  As a result  the elastic pole residue 
turns out to be also large  $|R|$=126 MeV. We note, that the extraction of the properties of  $P_{11}(1440)$
in the complex energy plane might require  a proper treatment of the 
$P_{11}(1440)\to\pi\Delta(1232)\to 2\pi N$ isobar decay
channel where the overlap of the self-energies of the $P_{11}(1440)$ and $\Delta(1232)$ states might be  important for
the determination of the  properties of $P_{11}(1440)$. This question will be addressed in \cite{shklyar:2012}.

As we already mentioned the results for   $P_{11}(1710)$ are controversial.
We find  159 MeV for the pole width. Somewhat greater value of 189 MeV  has been obtained 
in \cite{Tiator:2010rp,Batinic:2010zz}.
 The recent issue of PDG \cite{pdg} summarizes results 
for  the pole parameters taken  from 
four different analyses. 
Whereas the calculations \cite{Anisovich:2011fc,Hoehler:1993} give 200 MeV for the pole width,  Cutkosky  
obtains  a significantly lower value $\Gamma$=80 MeV \cite{Cutkosky:1990zh,Cutkosky:1979fy}. This results in a 
large spread of the resonance width given by PDG, see Table  \ref{poles}. 
The elastic residue is found to be small which is in accordance with the small decay
branching ratio to  $\pi N$. The similar conclusion has also  been  drawn in \cite{Tiator:2010rp}.

Investigation of the  $P_{13}$- wave  inelasticity \cite{Arndt:2006bf} shows that the $P_{13}(1720)$ state could  
have a strong decay flux into the $3\pi N$ channel \cite{Manley:1984}. 
Therefore the calculation of its pole width  might be affected by 
deficiencies in description of this channel. PDG estimations are based on several studies
where   $\Gamma_{\rm pole}= 120\pm 40$ by Cutkosky
\cite{Cutkosky:1979fy} is the lower limit.  The upper bound  $\Gamma_{\rm pole}=450\pm 100$\,MeV is given by the recent Bonn-Gatchina 
analysis \cite{Anisovich:2011fc}. Neither of these calculations includes the $3\pi N$ channel explicitely.

The situation with the second $P_{13}(1900)$ state is even more complicated. This resonance is rated by two stars 
in PDG and supposed to be rather broad.  The latest GWU  analysis \cite{Arndt:2006bf}  does not find any 
indication for this state. The present information about the pole parameters in PDG is based 
solely on the result of the Bonn-Gatchina calculations \cite{Anisovich:2011fc}
which deduce the pole mass  $1900\pm 30$\,MeV and the pole width  $200^{+100}_{-60}$\,MeV.
These values are very close to those derived in the present work.

The pole width of the $D_{13}(1520)$ state ( 94 MeV) turns out to be 10 MeV less than the
lower limit given in PDG\cite{pdg}.  The similar value of $\Gamma_{\rm pole}=$ 95\,MeV has also  been 
obtained in the J\"ulich model \cite{Doring:2009yv}. Some analyses find  additional poles associated with the $D_{13}(1700)$ 
and $D_{13}(1875)$ states \cite{pdg}. We do not  find any indication for $D_{13}(1700)$. The pole  
position for the second resonance is close  to the results of other calculation \cite{pdg}.  

Though the elastic residues for the $D_{15}(1675)$ and $F_{15}(1680)$  states are comparable 
 with the values given in PDG their
pole widths are somewhat lower than those obtained in other  studies \cite{pdg}. We also find an indication for the 
second state N(2000) with the pole  mass of 1900 MeV and  the width of $123$\,MeV, see Table\,\ref{poles}. 
This resonance has a small
coupling to the $\pi N$  final state what is in agreement with results from other calculations.

\begin{table}
 \begin{tabular}{ccccccc|cc}
 \hline
 \hline
 $N^*$  & mass (MeV) & $\Gamma_{tot}(MeV)$ &
 R$_{\pi N}$ & R$_{2\pi N}$ & R$_{\eta N}$
  & R$_{\omega N}$& $A^{p}_{\frac{1}{2}}$  &    $A^{p}_{\frac{3}{2}}$ 
  \\ 
 \hline
 \hline
   $S_{11}$(1535)
& 1526(2)  & 131(12) & 35(3)  & $ 8(2)$ & $58(4)$     & ---    &   91(4)  &     ---  \\                       
& 1526     & 136     & 34.4   & $ 9.5$  & $56.1 $     & ---    &   92     &  ---     \\ 
& 1536(10) & 150(25) & 45(10) & $ 5(5)$ & $42(10)$    & ---    &   90(30) &   ---  \\

   $S_{11}$(1650)                                                                           
& 1665(2)  & 147(14) & 74(3)  & $23(2)$ & $ 1(2)$     & ---  &   63(6)    &     ---     \\
& 1664     & 131     & 72.4   & $23.1 $ & $ 1.4 $     & ---  &   57       &  ---      \\
& 1657(13) & 150(30) & 70(20) & $15(5)$ & $ 10(5)$    & ---  &   53(16)   &     ---  \\
                                                                                            
 \hline                                                                                     
  $P_{11}$(1440)                                                                            
& 1515(15) & 605(90) & 56(2)  & $44(2)$ &   ---      & ---  &  -85(3)  &     ---  \\
& 1517     & 608     & 56.0   & $44.0$  &   ---      & ---  &  -84     &    --- \\
& 1445(25) & 300(150)& 65(10) & $35(5)$ &   ---      & ---  &  -60(4)  &    ---  \\

  $P_{11}$(1710)                                                                            
& 1737(17) &368(120)& 2(2)   & $49(3)$  & $45(4)$     &  3(2) &   -50(1)   &     ---  \\
& 1723     & 408    &  1.7   & $49.8 $  & $43.0$      &  0.2  &    -50    &     ---  \\
& 1710(30) &150(100)& 13(7)  & $65(25)$ & $20(10)$    &  13(2)&   24(10)  &     ---  \\
                                                                                           
 \hline                                                                                    
   $P_{13}$(1720)      
& 1700(10) & 152(2)   & 17(2) & $79(2)$ & $ 0(1)$    &  --- &  -65(2)   &     35(2)  \\
& 1700     & 152      & 17.1  & $78.7$  & $ 0.2$     &  --- &  -65      &   35       \\                  
& 1725(24) & 225(125) & 11(3) & $>70$   & $ 4(1)$    &  --- &   50(60)  &   -19(20)  \\ 

   $P_{13}$(1900)                                                                          
& 1998(3) & 359(10) & 25(1) & $61(2)$ & $ 2(2)$    & 10(3)&   -8(1)     &    0(1)     \\                                
& 1998    & 404     & 22.2  & $59.4 $ & $ 2.5$     & 14.9  &   -8       &    0         \\
& 1900(-) & 250(-)  & 10(-) &   ---   & $ 12(-)$    & 39(-)&   26(15)   &    -65(30)   \\                                  
                                                                                                                               
 \hline                                                                                
   $D_{13}$(1520)                                                                                                              
& 1505(4) &  100(2)  & 57(2) & $44(2)$ & $ 0(1)$    & ---  &  -15(1)    &    146(1)  \\
& 1505    & 100   & 56.6     & $43.4 $ &   1.2      &  --- &  -13       &  145        \\
& 1520(5) &  112(12)  & 60(5) & $25(5)$ &  2.3$\pm 10^{-3}$    & ---  &  -24(8)    &    150(15) \\
                                                                                      
   $D_{13}$(1875)                                                                            
& 1934(10) &857(100)& 11(1) & $69(2)$  & $ 0(1)$      & 20(5) &   11(1) &     26(1)\\
& 1934     & 859    & 10.5  & $68.7 $  & $ 0.5$       & 20.1  &   11     & 26      \\
& 1875(45) &220(100)& 12(10)& $70(20)$ & $ 3.5(3.5)$  & 21(7) &   18(10) &  -9(5)  \\
                                                                                                                       
 \hline                                                                                       
   $D_{15}$(1675)                                                                                                              
& 1666(2) & 148(1)  & 41(1) & $58(1)$ & $ 0(1)$  &  --- &    9(1)  &     21(1) \\
& 1666    & 148     & 41.1  & $58.5$  & $ 0.3 $  & ---  &    9     &    20      \\
& 1675(5) & 150(15) & 40(5) & $55(5)$ & $ 0(1)$  &  --- &    19(8) &     15(9) \\
                                                                                              
 \hline                                                                                                                        
   $F_{15}$(1680)                                                                             
& 1676(2) & 115(1) & 68(1) & $32(1)$ & $ 0(1)$  &  --- &    3(1)   &    116(1) \\
& 1676    & 115    & 68.3  & $31.6$  & $ 0.0 $  & ---  &    3      &    115     \\
& 1685(5) & 130(10)& 67(3) & $35(5)$ & $ 0(1)$  &  --- &    -15(6) &    132(13) \\
 
   $F_{15}$(2000)
& 1946(4)   & 198(2)   &  10(1) & $87(1)$ & $ 2(2)$  &  1(1) &   11(1) &     25(1)   \\
& 1946      & 198      &  9.9   & $87.2 $ & $ 2.0 $  &  0.4  &    10   &    25        \\
& 2050(100) & 350(200) &  15(7) & ---     &  ---     &  ---  &  35(15) &    50(14)  \\
 
 \hline
   \end{tabular}\\
\caption{Resonance parameters extracted  in the present study.  The uncertainties are given in brackets.
  Helicity decay amplitudes   are given in $10^{-3}$GeV$^{-\frac{1}{2}}$. 
1st line: present study; 2nd line: \cite{shklyar:2004b}, 3th line: \cite{pdg}.
  (-): the validity range is not given.
 \label{res_param}}
\end{table}

\begin{table}
 \begin{tabular}{ccccc}
 \hline
 \hline
             &  Re\,$z_0$(GeV)   &  -2Im\,$z_0$(MeV)  & $|$R$|$(MeV) & $\theta^0$   \\
 \hline
 \hline

$S_{11}(1535)$  &      1.49       &    100     &       15    &      -51       \\ 
                &    1.49-1.53    &  90-250    &    30...70  &     -1...-30   \\ 
 \hline
$S_{11}(1650)$  &      1.65       &     89     &       19    &      -46       \\     
                &      1.64-1.67  &  100-170   &   20-50     &    -50...-80   \\    
 \hline
 \hline
$P_{11}(1440)$  &      1.386      &     277    &      126    &      -60       \\     
                &   1.35-1.38     &  160-220   &      40-52  &    -75...-100   \\     
 \hline
$P_{11}(1710)$  &      1.67       &    159     &       11    &        9         \\     
                &    1.67-1.77    &  80-380    &     2-15    &        -160...+190   \\     
 \hline
 \hline
$P_{13}(1720)$  &      1.67      &    118     &       12    &      -45          \\     
                &    1.66-1.69    &  150-400   &       7-23  &  -90...-160     \\     
 \hline

$P_{13}(1900)$ &    1.91         &    173     &       10    &      -64   \\     
               &    1.870-1.93    &  140-300   &       1-5   &    45...-25   \\     
 \hline
 \hline
$D_{13}(1520)$ &   1.492          &     94     &           27&      -35   \\
               &    1.505-1.515   &    105-120 &     32-38   &      -5...-15   \\
 \hline     
$D_{13}(1875)$ &      1.81       &     98     &        3    &      -76          \\     
               &      1.8-1.95    &  150-250   &        2-10 &   20...180  \\     
 \hline
 \hline
$D_{15}(1675)$ &      1.64        &    108     &       20    &      -49   \\     
               &   1.655-1.665    &   125-150  &     22-32   &  -21...40   \\     
 \hline
 \hline 
$D_{15}(1680)$ &      1.66         &     98    &       33  &      -32   \\     
               &      1.665-1.68  &   110-135 &      35-45  &   0...-30   \\     
 \hline
$F_{15}(2000)$ &      1.90        &    123  &       11  &       -6   \\   
               &     1.92-2.15     &  380-580  &    20-115  &       -60...-140   \\  
 \hline 
   \end{tabular}\\
\caption{Pole positions  and elastic pole residues. First line: present study,
second line: values from PDG \cite{pdg}.\label{poles}} 

\end{table}

\section{Results and discussion}
\label{results}
\label{pe}
The lack of the experimental data for the pion-induced reactions does not provide enough constraints on 
the resonance parameters. Also the discrepancy among various measurements (see Section~\refe{data}) does 
not allow  for a  consistent description of the data in a full kinematical region. 
 While the contribution from the $S_{11}(1535)$ state is well established 
the reaction dynamics  above W=1.6 GeV is still under discussion. 
One of the early Giessen coupled-channel calculations  \cite{Penner:2002a,Penner:2002b,Shklyar:2006xw} 
found  a  destructive interference between $S_{11}(1535)$  and $S_{11}(1650)$ states. 
The second  suggestion  is a strong contribution from the  
$P_{11}(1710)$-resonance excitation above W=1.68 GeV. This resonance was established in the early 
single-channel Karlsruhe-Helsinki and Carnegie Mellon-Berkeley analyses (see PDG \cite{pdg} and references 
therein). The  independent study of the $\pi N \to (\pi/\eta) N$ reactions by the  Zagreb group 
\cite{Ceci:2006ra}  provides an additional evidence for the existence of  $P_{11}(1710)$. The result 
of \cite{Ceci:2006ra} confirm the assumption made in \cite{Penner:2002a,Penner:2002b} on 
the important contribution from this state to the $\eta$-production.
 However the recent analysis from the GWU  group \cite{Arndt:2006bf} finds  no evidence 
for this state. The absence of a clear signal in the $P_{11}$ partial wave of the elastic $\pi N$  scattering
does not necessarily   mean that this state does not exist. If the coupling to the 
final $\pi N$ state is small, the effect from this state  might not be seen in $\pi N$ scattering.
 The evidence  for the signal from the  $P_{11}(1710)$ resonance has also been reported from the
study of the $\pi N \to K\Lambda$ reaction \cite{Ceci:2005vf}. 
 On the other hand the result of the Bayestian analysis performed by the 
Gent group \cite{DeCruz:2012bv} demonstrates that $P_{11}(1710) $
is not needed to describe the  $K\Lambda$ photoproduction. An opposite conclusion was drawn by  the
 Bonn-Gatchina group which finds  
decay branching ratio of $23\pm 7$\%  of this state to $K\Lambda$ \cite{Anisovich:2011fc}.

\begin{figure}
  \begin{center}
    \includegraphics[width=8cm]{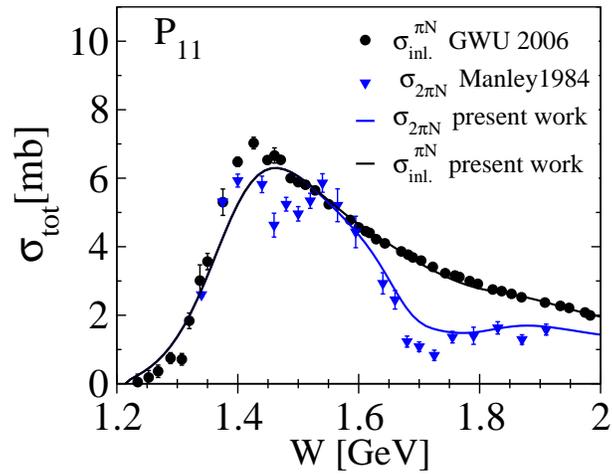}
    \caption{(Color online) Calculated $\pi N$ inelasticity and $\pi N\to 2\pi N$ cross section in the $P_{11}$ partial wave
     in comparison with the results from \cite{Arndt:2006bf} (GWU\,2006) and \cite{Manley:1984}(Manley\,1984).  
      \label{2piN}}
  \end{center}
\end{figure}

Another indication for this state comes from the analysis of an inelastic flux in the $P_{11}$ partial wave.
In Fig.~\refe{2piN} the total inelasticity from the GWU analysis vs. the total $2\pi$ cross section extracted
in \cite{Manley:1984} is compared. The difference between the total $\pi N$ inelasticity and 
the total $2\pi N$ cross section at W=1.7 GeV in the $P_{11}$-wave
can be attributed to the sum of inelastic  channels like $3\pi N$,  $\eta  N$, $\eta \pi N$ etc. We assume here 
that the observed  difference is  due to the $\eta N$ production channel dominated by the $P_{11}(1710)$ state.
As $g_{\pi N N^*(1710)}$ is assumed to be small this raises the question about the magnitude of the  
$P_{11}(1710)$ contribution  in the $\pi N \to \eta N$  reaction.
However the  situation in $\eta$-production is different from the $\pi N$ elastic scattering. Here the 
contribution from  $P_{11}(1710)$ is proportional to the product $g_{\pi N N^*(1710)}g_{\eta N N^*(1710)}$, 
where $g_{\pi N N^*(1710)}$ is the coupling constant at  the $N(1710)\to \eta N$ transition vertex. 
It follows  that the contribution  from the $P_{11}(1710)$ can be significant provided that  
$g_{\eta N N^*(1710)}$ is large enough. The interplay with background and coupled-channel rescattering would 
further increase this effect.
   
\subsection{$\pi N \to \eta N$}
\begin{figure}
  \begin{center}
    \includegraphics[width=18cm]{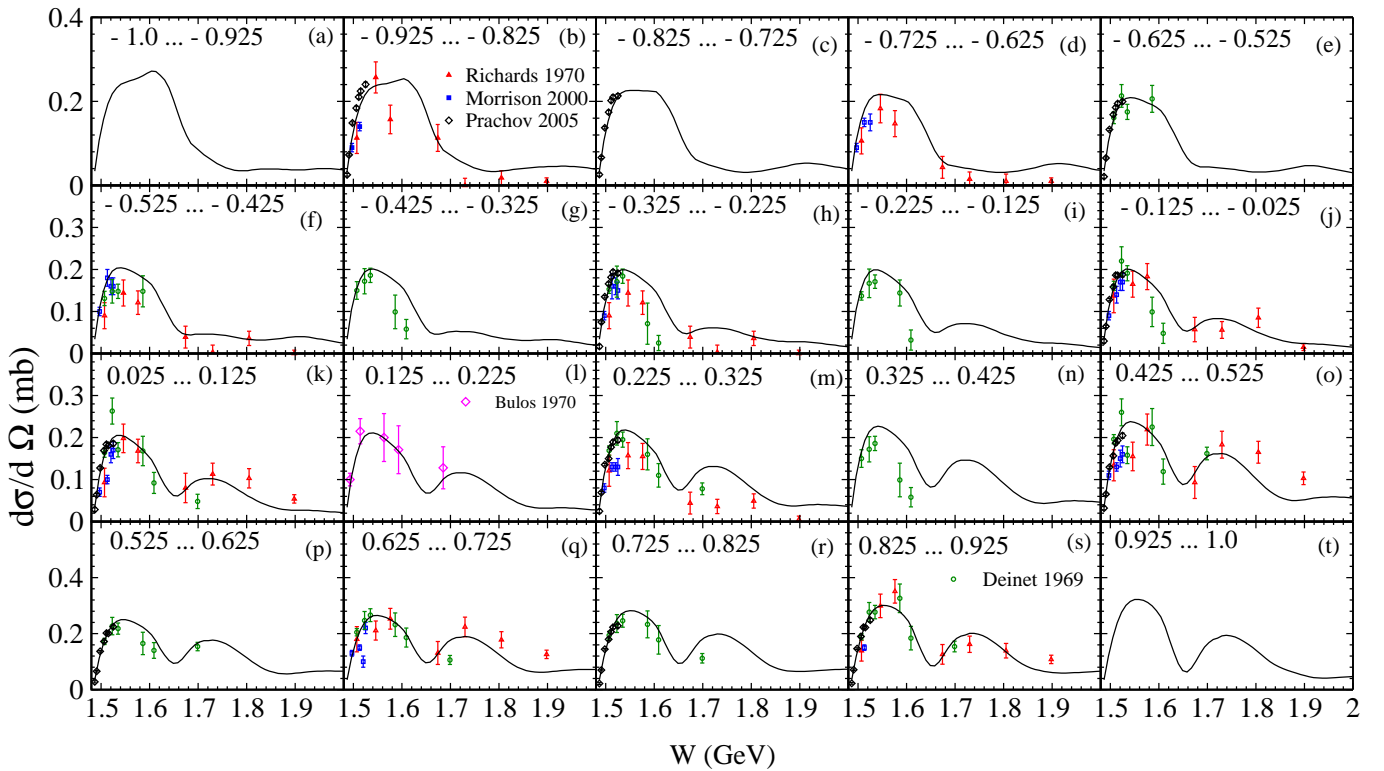}
    \caption{(Color online) Calculated differential  $\pi^- p \to \eta n$  cross section in comparison with the 
 experimental data from:  Prakhov\,2005:\cite{Prakhov}, Deinet\,1969:\cite{Deinet:1969cd},
Richards\,1970:\cite{Richards:1970cy}, Morrison\,2000:\cite{Morrison:2000kx}.
      \label{pe_dif}}
  \end{center}
\end{figure}

The results of our calculations are presented in Fig.~\refe{pe_dif} in comparison with the world data. 
The first peak at W=1.54 GeV is related to the well established $S_{11}(1535)$ resonance contribution.
Though the effect from the $S_{11}(1650)$ state is hardly visible in the differential cross section
this state plays an important role leading to the destructive interference between  $S_{11}(1535)$
and $S_{11}(1650)$  as it has been pointed out in our previous  calculations
\cite{Penner:2002a,Penner:2002b}.

The second rise is due to the $P_{11}(1710)$ resonance. This state has a small branching 
ratio to the $\pi N$  system but due to the large  $\eta$-coupling this resonance affects the 
production cross section at W=1.7 GeV. The coupled-channel effects  and interference
with other partial waves further enlarge  the overall contribution from  this state. 

\begin{figure}
  \begin{center}
    \includegraphics[width=10cm]{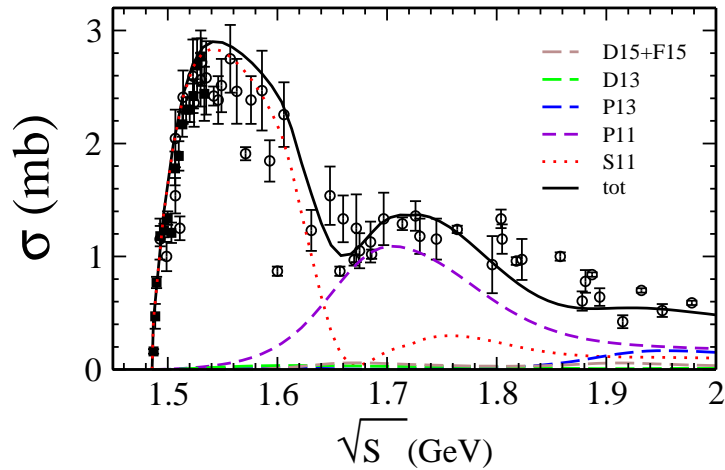}
    \caption{(Color online) Total partial wave cross section  $\pi^- p \to \eta n$  vs. experimental data.
      \label{pe_tot}}
  \end{center}
\end{figure}

The total partial wave cross sections are shown in Fig.~\ref{pe_tot}. The destructive interference 
between the $S_{11}(1535)$
and $S_{11}(1650)$ leads to the dip in the total $S_{11}$-partial wave cross section around W=1.64 GeV 
(dotted line). The effect from the $P_{11}(1710)$ state is shown by the dashed line, Fig.~\ref{pe_tot}.
The contributions from other partial waves are found to be small.  We also corroborate our previous 
results \cite{shklyar:2004a} where only minor contributions from spin $J\ge\fth$ resonance states 
were obtained. Both  t-channel  $a_{0}$ and $\rho$ meson exchange and $u-$channel graphs give  
small effects.  The inclusion of the higher spin state $D_{13}(1520)$  into the calculations 
is still important to reproduce the correct shape of the cross section. 
This feature is also found in many other calculations, e.g.\cite{Penner:2002a,Batinic:1995}.  
It is interesting to note that importance of the $P_{11}(1710)$ resonance 
contribution has recently  been found in \cite{Shrestha:2012va} which is  in line with our previous 
results \cite{Shklyar:2006xw,Penner:2002b}.

Since the main contributions  in our calculations come mainly from the $S_{11}$ and $P_{11} $ partial waves 
it is interesting to trace back the interference effect between them. Neglecting the higher partial waves
the  differential cross section can be written in the form
\bea
\frac{d\sigma}{d\cos\,(\theta)}~\sim~ 1+ \alpha \,\sin^2\left(\frac{\theta}{2}\right), 
\label{s_p_interf}
\eea
where $\theta$ is a scattering angle and $\alpha =\left(\frac{|S_{11} - P_{11}|^2}{ |S_{11} + P_{11}|^2}-1 \right) $
only depends 
on the c.m. energy. Then the angular distribution  should have a maximum
(minimum) at forward angles depending on the relative phase between the nonvanishing  $S_{11}$ and   $P_{11}$ 
amplitudes. In our calculation the interference between $S_{11}$ and $P_{11}$ partial waves produces
a peak  at forward scattering angles and energies  above W=1.67 GeV, see Fig.~\refe{pe_dif}. 
As a result the signal from the $P_{11}(1710)$ resonance becomes more transparent for  forward 
scattering.  
This is in line with the data of Richards et al \cite{Richards:1970cy} 
confirming our guess about the  production mechanism. The inclusion of higher partial waves would  
modify Eq.~\refe{s_p_interf}. However  these contributions  are relatively
small  (see Fig.\refe{pe_tot}) thus producing   only minor  deviations from the distribution 
Eq.~\refe{s_p_interf}.
 
Note, that due to numerous problems with the experimental data  our calculations above W=1.6 GeV 
are only partly  constrained by experiment. Indeed, once the data \cite{Baker:1979aw,Brown:1979ii} 
are neglected there are only 30 datapoints from experiment \cite{Richards:1970cy}.  This data has  
relatively large error bars and  seems not to be fully consistent with other measurements 
\cite{Clajus:1992dh}. Therefore, the results for the differential cross section 
might be regarded as a prediction rather than an outcome of the fit.
 This demonstrates an  urgent need for new  measurements of the 
$\pi^- N \to \eta N$ reactions above W=1.6 GeV.  This would be a challenge for the the  
upcoming pion-beams experiment carried  out by the HADES collaboration at GSI.

\subsection{$\eta N \to \eta N$  amplitude  and $\eta N$ scattering lengths}
\begin{figure}
  \begin{center}
    \includegraphics[width=7cm]{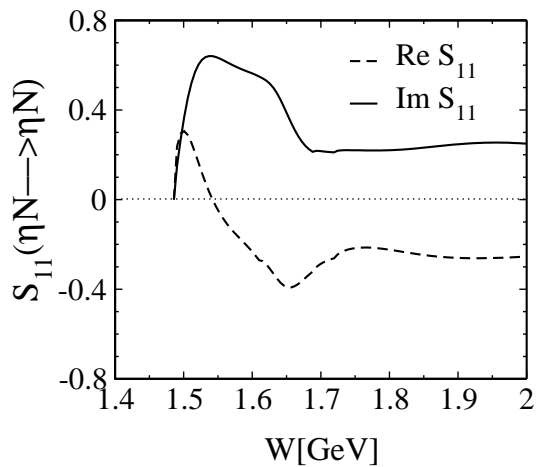}
    \caption{Calculated $S_{11}$ partial wave amplitude  of the elastic  $\eta N$ scattering.
      \label{ee_S11}}
  \end{center}
\end{figure}

The result for the $\eta N \to \eta N$ transition  amplitude in the $S_{11}$ partial wave  is presented in 
Fig.~\ref{ee_S11}. Close to threshold the  elastic $\eta N$ scattering  is completely determined by the 
contribution of the $S_{11}(1535)$ resonance. At higher energies the excitation of  $S_{11}(1650)$
also becomes   important. The interference between those two $S_{11}$-states produces an excess structure in the imaginary 
part of the  amplitude at W=1.65 GeV.

The rapid variation of the $S_{11}$-amplitude close to threshold indicates that this energy  dependence  should be 
taken into account when the  $\eta N$ scattering length is calculated. 
Here we use the definition  for the effective range expansion from \cite{PhysRevC.55.R2167}:
\bea
\frac{q_{\rm c.m.}}{ S_{11}^{\,\eta N}} + i q_{\rm c.m.} = \frac{1}{a_{\eta N}} + \frac{r_0}{2} q_{\rm c.m.}^2 +s\, q^4_{\rm c.m.},
\eea
where $S_{11}^{\,\eta N}$ is an elastic partial S-wave amplitude, and $a_{\eta N}$, $r_0$ and $s$ are scattering length, effective range,
and effective volume respectively. The results are shown in Table~\ref{tscatt_lengths} in comparison with 
values deduced from other  coupled-channel calculations ( results published before 1997 are discussed in
\cite{PhysRevC.55.R2167} ). The obtained value of $a_{\eta N}$ is very close  to  our previous results 
\cite{Penner:2002a}.    The values for  the real part  deduced in \cite{Batinic:1996me} and  \cite{PhysRevC.55.R2167}
are lower than in this work. The study \cite{Batinic:1996me} gives  1.550 GeV for the mass and  204 MeV 
for the width of the $S_{11}(1535)$ state which are somewhat greater than in the present calculation.
This could be one of the reasons   for the  differences in   $Re\,a_{\eta N}$. 

In \cite{PhysRevC.55.R2167}  only the $S_{11}(1535)$ state is taken  into account 
to calculate transition amplitudes to the $\eta N$ channel. 
Since the  parameters of $S_{11}(1535)$ in \cite{PhysRevC.55.R2167}  are  close to the values obtained in the present study 
the observed difference in ${\rm Re}\,a_{\eta N}$ might be attributed to the different treatment of  background 
contributions which have  been assumed  in \cite{PhysRevC.55.R2167} to be energy-independent. The second piece of uncertainty
is related to the quality of the world data of   $\pi N \to \eta N$ scattering. Hence, precise measurements of this 
reaction would provide an additional constraint on  $\eta N$ scattering length.

The non-vanishing imaginary part of  $ a_{\eta N}$  is mostly driven by  rescattering in the $\pi N$ channel. Since the 
largest contributions to the scattering length are produced by the $S_{11}(1535)$ state the imaginary part  of $a_{\eta N}$ is  
strongly influenced  by the decay branching ratio of this resonance  to $\pi N$. Only a minor effect is found  from 
the rescattering induced by background contributions and  inelastic flux to  the $2\pi N$ channel. 
Since the $\pi NN^*(1535)$ coupling is  well fixed 
an agreement in ${\rm Im}( a_{\eta N})$ between various model calculations can be expected provided that unitarity 
is maintained. 

The obtained value of the  scattering length should be taken with care when in-medium  properties  of the $\eta$-meson 
are considered. As it has already been  pointed out in  \cite{PhysRevC.55.R2167} the $S_{11}$ amplitude has a strong energy 
dependence - a  feature which might affect the $\eta$-potential. The second reason is that  properties of 
the $S_{11}(1535)$ resonance might also be  subjected to in-medium modifications \cite{Lehr:2003km}. Both effects should be taken into 
account when  $\eta$-meson properties in nuclei are studied.

 \begin{table}
  \begin{center}
    \begin{tabular}
      {l|c|c }
      \hhline{===}
     Reference & $a_{\eta N}$(fm) & $r_0$(fm)       \\ 
      \hhline{===}
      present work             &   0.99$\pm $0.08 + i0.25$\pm$0.06   & -1.98$\pm$0.1 - i0.43$\pm$0.15 \\
      \cite{Penner:2002a}      & 0.99            + i0.34              & -2.08 - i0.81 \\

       \cite{Batinic:1996me}   & 0.734$\pm$0.026 + i0.269$\pm0.019$       &                   \\
       \cite{PhysRevC.55.R2167}&  0.75$\pm 0.04$ + i0.27$\pm 0.03$    & -1.5$\pm$0.13 - i0.24$\pm$0.04 \\
       \cite{Lutz:2001}        &    0.43+ i0.21                       &                              \\
      \hhline{===}
    \end{tabular}
  \end{center}
  \caption{ Calculated scattering length and effective range in comparison with results from other works.
    \label{tscatt_lengths}} 
\end{table}

\subsection{$\gamma N \to \eta N$ below 1.89 GeV}
\begin{figure}
  \begin{center}
    \includegraphics[width=10cm]{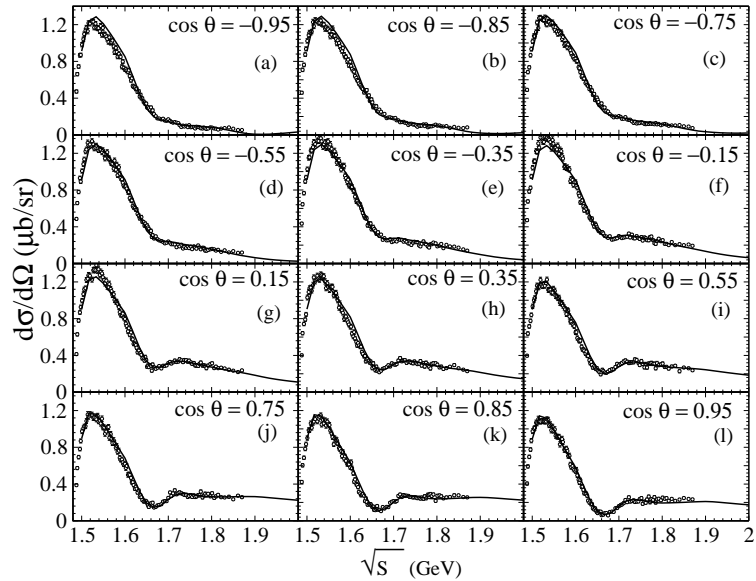}
    \caption{ Differential $\eta p$  cross section vs. recent MAMI data \cite{McNicoll:2010qk}.
      \label{ge_dif}}
  \end{center}
\end{figure}

The results of our calculation of the   differential cross section in comparison with the recent
Crystal Ball/MAMI
measurements are shown in Fig.~\refe{ge_dif}. Our calculations demonstrate a nice agreement with
the experimental data  in the whole kinematical region. The first peak is related to the $S_{11}(1535)$
resonance contribution. Similar to the $\pi^- p \to \eta n$ reaction the $S_{11}(1650)$ and $S_{11}(1650)$  
states interfere destructively producing a dip around W=1.68 GeV.
 Though the  effect from the  $P_{11}(1710)$ 
state is only  minor,  the contribution from this resonance produces a rapid change in the $M_{1-}$
photoproduction multipole, see Section \ref{multipoles}.  
The coherent sum of all partial waves leads to the more pronounced effect from the dip 
at  forward angles.
Note that 
the  resonance contribution to the photoproduction reaction  stems from two  sources: the first is related 
to the direct electromagnetic excitation of the nucleon resonance and the second   comes from 
rescattering e.g. $\gamma p \to \pi N \to \eta N$,  Eq.~\refe{photo}. At this stage the hadronic 
transition amplitudes  e.g. $T_{\pi N \to\eta N}$ become an important part of the production mechanism. 
The sum of these contributions in the $P_{11}$ wave turns out to be destructive  which reduces the overall contribution 
from the $P_{11}(1710)$ state.
We also corroborate our previous findings \cite{Shklyar:2006xw} where a  small  
effect from the $\omega N$ threshold  was found.   
We also do not find any strong indication for  contributions from  a hypothetic  narrow $P_{11}$ 
state with a  width of 15-20 MeV around  W=1.68 GeV. 
It is natural to assume that the contribution from this state would induce a strong modification 
of the beam asymmetry for energies close  to the mass of this state. This is because the beam asymmetry
is less sensitive to the absolute magnitude of the various partial wave contributions but strongly affected 
by the relative phases between different partial waves. Thus even a small admixture of a  contribution
from a narrow state might result into a strong modification of the beam asymmetry in the energy 
region of  W=1.68 GeV.

\begin{figure}
  \begin{center}
    \includegraphics[width=10cm]{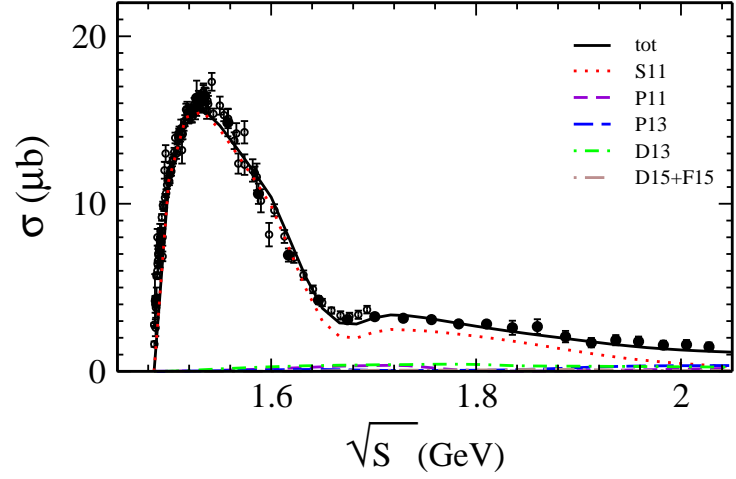}
    \caption{(Color online) $\gamma p \to \eta p$ partial wave cross sections vs. measurements \cite{Crede:2003ax,Crede:2009zzb,Bartholomy:2007zz}.
      \label{ge_tot}}
  \end{center}
\end{figure}

\begin{figure}
  \begin{center}
    \includegraphics[width=12cm]{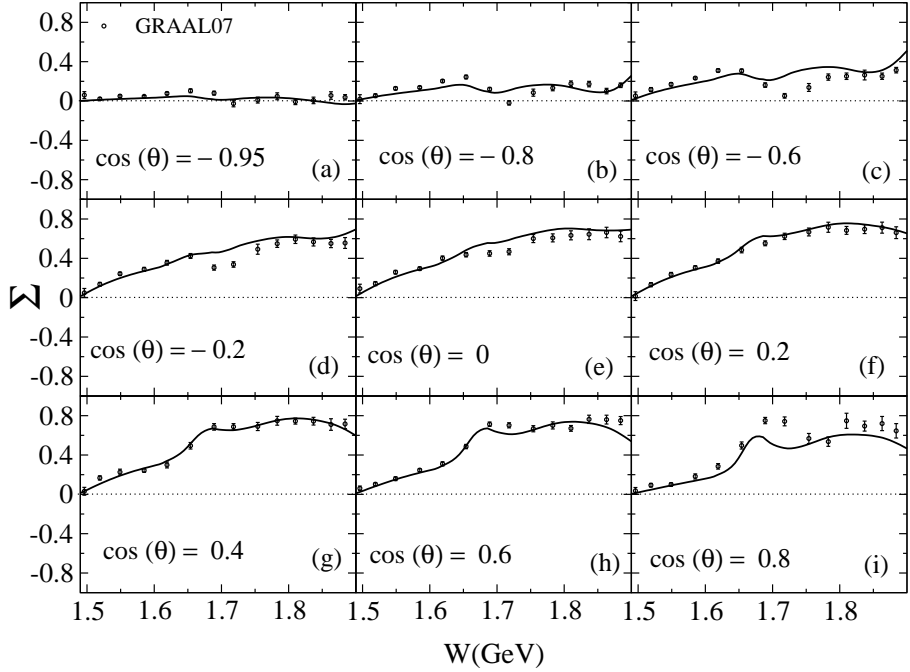}
    \caption{Calculated beam asymmetry. Experimental data  are taken from \cite{Bartalini:2007fg}(GRAAL07).
      \label{ge_sig}}
  \end{center}
\end{figure}

In Fig. \refe{ge_sig} we show the calculation of the photon-beam asymmetry in comparison with the GRAAL 
measurements \cite{Bartalini:2007fg}. One can see that even close to the $\eta N$ threshold where our calculations 
exhibit a dominant $S_{11}$ production mechanism (see Fig.~\refe{ge_tot} ) the beam asymmetry is 
nonvanishing for angles  $\cos(\theta)\ge-0.2$. This shows  
that this observable is very sensitive to  very small contributions from   higher partial waves.
At W=1.68 GeV  and forward angles the GRAAL measurements show a rapid change of the asymmetry 
behavior. We explain this effect by a destructive interference between the  $S_{11}(1535)$ and $S_{11}(1650)$
resonances which induces the  dip at W=1.68 GeV in the $S_{11}$ partial wave. The strong drop in  
the $S_{11}$ partial wave  modifies the interference between $S_{11}$ and other partial waves and  
changes the asymmetry behavior.
Note that the interference between $S_{11}(1535)$ and $S_{11}(1650)$ and 
the interference between different partial waves are of different nature.
 The overlapping of the
 $S_{11}(1535)$ and  $S_{11}(1650)$ resonances does not simply mean a coherent sum of two independent 
contributions, but also includes   rescattering (coupled-channel effects). Such interplay is hard to simulate
by the simple sum of two Breit-Wigner forms since it does not take into account 
rescattering due to the coupled-channel treatment.

The GRAAL collaboration finds   no evidence for a narrow state around W=1.68 GeV. 
We also find no strong need  for the narrow $P_{11}$ resonance contribution to describe the 
asymmetry data.  Taking  contributions from the  established states into account our results are in 
close agreement with the experimental data \cite{Bartalini:2007fg}.

\subsection{$\gamma N \to \eta N$  above 1.89 GeV}

\begin{figure}
  \begin{center}
    \includegraphics[width=14cm]{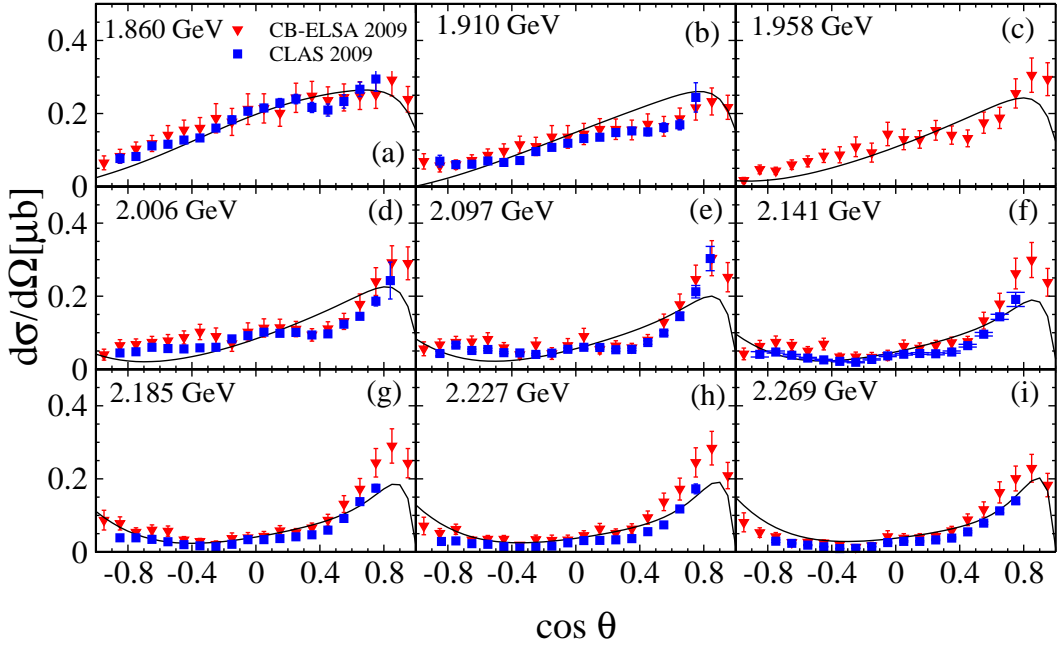}
    \caption{(Color online) Differential $\eta p$ cross section as a function of the scattering angle.
       The data are taken from CLAS\,2009:\cite{Williams:2009yj} and  CB-ELSA:\cite{Crede:2009zzb}. \label{ge_dif2}}
  \end{center}
\end{figure}

\begin{figure}
  \begin{center}
    \includegraphics[width=16cm]{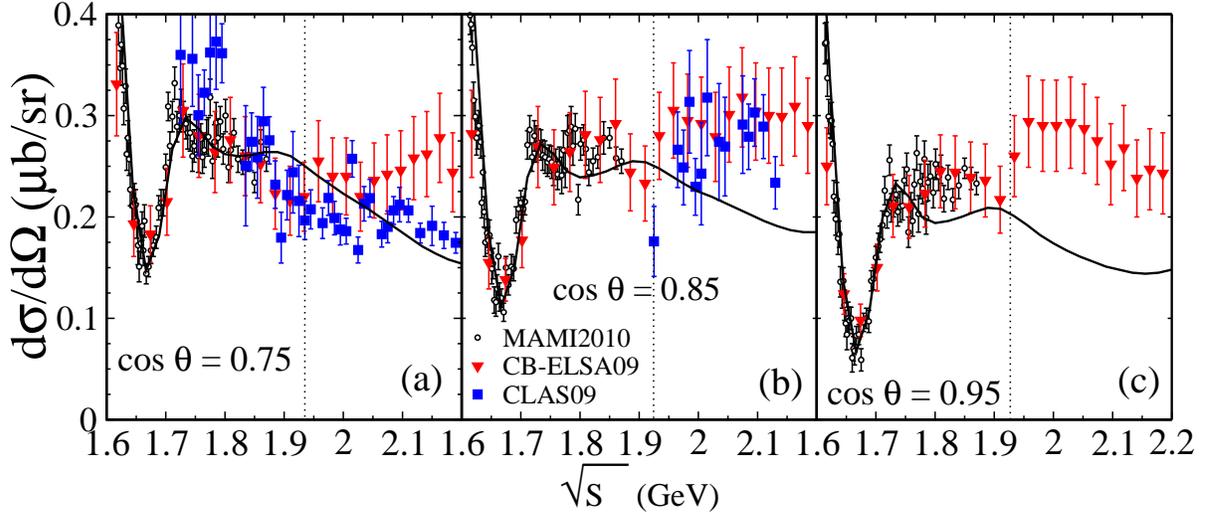}
    \caption{(Color online) Differential $\eta p$ cross section as a function of c.m. energy at fixed forward angles.
             Data are taken from CLAS\,2009:\cite{Williams:2009yj}, CB-ELSA:\cite{Crede:2009zzb}, and MAMI2010:\cite{McNicoll:2010qk}.
      \label{ge_d3f}}
  \end{center}
\end{figure}

Since the  MAMI measurements are available up to W=1.89 GeV the calculations in the region  W=1.89 ...2.GeV
are constrained by the combined  data set constructed out of the recent CLAS and CB-ELSA/TAPS
 \cite{Williams:2009yj,Crede:2009zzb} data.  Due to some  inconsistencies between these 
two experiments \cite{Dey:2011rh,Sibirtsev:2010yj}  we did not try to fit the data  above W=2.GeV but instead 
extrapolate our calculation into the higher energies. In this region the $t$-channel exchange starts to 
play a dominant role. One of the accepted prescriptions is to use a  Reggeized $t$-channel meson  exchange  
as suggested in \cite{Chiang:2002vq}. We do not follow this procedure here but include all $t$-channel
exchanges into the interaction kernel. This allows for a  consistent unitary  treatment of resonance and background
contributions. The calculated differential cross section  is presented 
in Fig.~\refe{ge_dif2} as a function of the scattering angle. 
Except for  the energy bin W = 2.097 GeV  our results are found to be in  close 
agreement with the CLAS measurements. The major  contribution  to the differential cross section at forward
angles  comes from  $\rho$- and $\omega$-meson exchanges. The effect from the $\phi$-meson 
is small due to the weakness of the $\phi NN$ coupling as dictated by the OZI rule 
\cite{Okubo:1963fa,Zweig:1964jf,Iizuka:1966fk}.
 We also checked for the contributions from the Primakoff 
effect which is found to be negligible at these  energies.

It is interesting to compare our calculations with the data \cite{Williams:2009yj,Crede:2009zzb} at 
forward angles plotted as a function of the c.m. energy, see Fig.~\refe{ge_d3f}. 
The cusp due to the $\omega N$
production threshold is clearly seen in our calculations around W=1.72 GeV. The quality of the data is still 
not good  enough to unambiguously resolve the cusp induced by the $\omega N$ threshold in the experimental data. 
Note, that the calculations 
are done assuming a stable $\omega$-meson. Taking into account the final $\omega$-width would smear out this effect.
Since the $\omega N$ threshold lies 45 MeV above the dip position ( W=1.68 GeV) we conclude that this effect cannot explain
the dip in the differential cross section. This conclusion is opposite to that drawn 
in  \cite{Anisovich:2011ka}. 

The discrepancy between the CLAS \cite{Williams:2009yj} and CB-ELSA/TAPS data is better seen  at $\cos(\theta)=0.75$ 
whereas for $\cos(\theta)=0.85$ the measurements  are found to be in better agreement. 
One of the interesting features observed in the recent CB-ELSA data is a sudden rise of the differential cross 
section at W=1.92 GeV. The effect is more pronounced at $\cos(\theta)=0.85...0.95$ and  is absent at other scattering 
angles. This phenomena might be attributed to sidefeeding from  of one of the inelastic channels (e.g. $\phi N$, $a_0(980) N$, 
$f_0(980)$,  or $\eta' N$). However  the  problem with normalization inconsistencies 
between the CLAS and CB-ELSA data should be solved
first before any  physical interpretation can be given.

\subsection{eta-photoproduction multipoles  \label{multipoles}}
\begin{figure}
  \begin{center}
    \includegraphics[width=8cm]{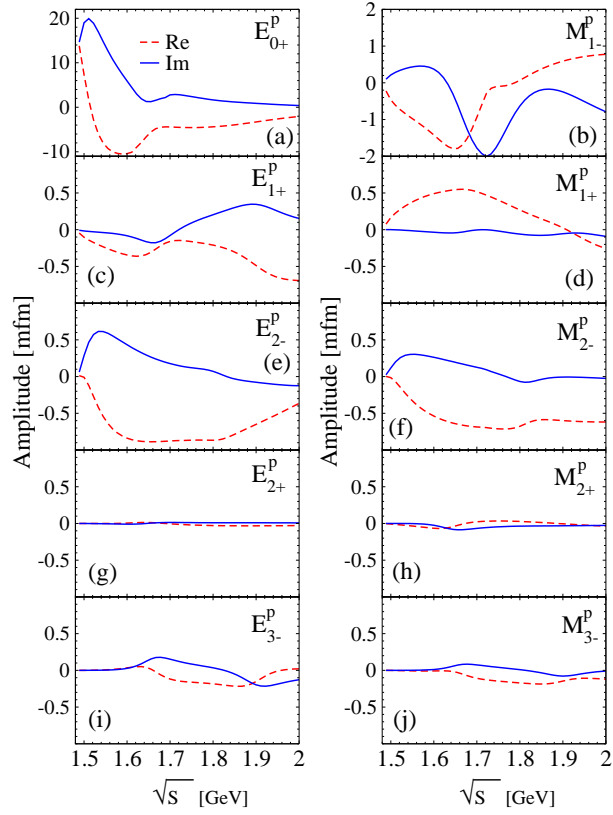}
    \caption{(Color online) $\gamma p \to \eta p$ multipoles extracted in the present study. \label{ge_multiploes}}
  \end{center}
\end{figure}

The extracted $\gamma p \to \eta p$ multipoles are presented in Fig.~\ref{ge_multiploes}.
The major contribution to the  $E_{0+}$ multipole comes from the $S_{11}(1535)$ resonance. 
The second $S_{11}(1650)$ plays an important role  in the region W=1.6...1.7 GeV. 
We corroborate our previous results \cite{Shklyar:2006xw} where only a small effect from the spin-$\ffh$
states has been found. A very small signal  from  the $F_{15}(1680)$ resonance is seen in 
the  $E_{3-}$ and  $M_{3-}$ amplitudes at W=1.68 GeV. 

It is interesting  to note that the effect of  $D_{13}(1520)$ is clearly seen in the  $E_{2-}$ and  $M_{2-}$ though
the overall contribution from this state  turns out to be   small. 
The $M_{1-}$ multipole is affected  by the Roper and   $P_{11}(1710)$ resonances leading to the rapid change in  
both real and imaginary parts of the amplitude at W=1.7 GeV. In the region W=1.48...1.6 GeV  both the imaginary 
and the real parts of all multipoles with $l\neq 0$ are of the order of magnitude smaller than  $E_{0+}$ 
due to the strong dominant contribution from $S_{11}(1535)$. However for higher energies the influence of  
amplitudes  with $l\neq 0$ becomes also important.

\section{Conclusion}
We have performed a coupled-channel analysis of pion- and photon-induced reactions including the recent 
eta-photoproduction data from the  Crystal Ball/MAMI collaboration.  In the region  W=1.89...2.0 GeV our solution is constrained by 
 the combined dataset built from  the  recent CLAS and CB-ELSA/TAPS measurements. The dip 
in the differential cross-sections at W=1.68 GeV  reported in \cite{McNicoll:2010qk}  is described in 
terms of an  interference of the $S_{11}(1535)$ and  $S_{11}(1650)$ states. 
We stress that such an interference  also includes coupled-channel effects and
rescattering which is hard to simulate by a simple sum of two 
Breit-Wigner contributions. 
The additional contribution at W=1.68 GeV comes from the $M_{1-}$ multipole where the excitation of the $P_{11}(1710)$
leads to a rapid change of the real and imaginary parts of the amplitude.
We conclude that the cusp due to the $\omega N$ threshold seen at 1.72 GeV is not important
for the explanation of the dip at W=1.68 GeV. However the quality 
of the data is still not sufficient to  resolve the threshold effect  completely.

Above W=1.9 GeV the $t$-channel $\rho$- and $\omega$-exchanges start
to play a dominant role in the calculations. The effect  from the $\phi$-meson exchange is less 
important because of the smallness of the $\phi NN$ coupling. We have also checked for the contribution from the 
Primakoff-effect which is found to be negligible. In the region W=1.9...2.2~GeV our calculations 
tend to be in closer agreement with the CLAS data.

It is interesting to note that above W=1.92 GeV the cross sections of the CB-ELSA/TAPS collaboration 
indicate a sudden rise from 0.2~$\mu$b up to 0.3~$\mu$b. The effect is observed only for scattering angles  
$\cos(\theta)=0.85...0.95$.   This phenomenon might be attributed to sidefeeding from of one of the inelastic 
channels (e.g. $\phi N$, $a_0(980) N$, $f_0(980)$,  or $\eta' N$). However the origin of 
the normalization discrepancies between the CLAS and CB-ELSA/TAPS data should first be understood 
before any  physical interpretation can be given.

In the $\pi^- p \to \eta n$ reaction the main effect comes from three resonances $S_{11}(1535)$,
$S_{11}(1650)$, and $P_{11}(1710)$. Similar to  eta-photoproduction on the proton the overlap
of the  $S_{11}(1535)$ and $S_{11}(1650)$ states produces a dip around W=1.68 GeV. For  energies  $W>1.68$ GeV
the contribution from  $P_{11}(1710)$ is found  to be important. 
The above reaction mechanism for the $(\gamma/\pi)N\to\eta N$ reaction is in line with our early 
findings \cite{Shklyar:2006xw} where the resonance like-structure in $\eta$-photoproduction at W=1.68 GeV 
on the neutron  was explained by the excitations of the $S_{11}(1650)$, and $P_{11}(1710)$
resonances.

We conclude that  further progress in understanding of  $\eta$-meson production
would be hardly possible without new measurements of the $\pi N\to \eta N$ reaction.  
The experimental investigation  of this reaction would help to establish   the resonance contributions
to the $\eta$-photoproduction above $W >1.6 $ GeV. Finally, the study of the $\eta N$-channel with 
pion beams  would solve the question whether the observed  phenomena in $\eta$ photoproduction  
have their counterparts in $\pi N \to\eta N$ scattering.

\bibliographystyle{h-physrev}
\bibliography{tau1}

\end{document}